\documentclass[journal]{article}                                  % Needed to meet printer requirements.

\usepackage{arxiv}
\usepackage{hyperref}
\usepackage{graphicx}          % Include this line if your 
\usepackage{amsmath} % assumes amsmath package installed
\usepackage{amssymb}  % assumes amsmath package installed

\usepackage{amsthm}
\usepackage{color,soul}
\usepackage{flushend}
\usepackage{cite}
\usepackage{mfirstuc}
\usepackage{subfig}
\usepackage{mathtools}
\usepackage{xcolor}
\usepackage{array,multirow}
\usepackage{algorithm} 
\usepackage{algorithmic}  
\usepackage[linesnumbered,algo2e,ruled,vlined,norelsize]{algorithm2e} 
\allowdisplaybreaks
\usepackage{tikz}
\usepackage{textcomp}
\usepackage{hyperref}
\usepackage{lipsum}

\usepackage{authblk}
\title{\LARGE \bf
Resilient Voltage Estimation for Battery Packs\\ Using Self-Learning Koopman Operator
}

\author[1]{Sanchita Ghosh}
\author[1]{Tanushree Roy}
\affil[1]{Department of  Mechanical  \& Aerospace Engineering  Engineering, Texas Tech University, Lubbock, TX 79409, US. Emails:~{\tt\small sancghos@ttu.edu, tanushree.roy@ttu.edu}.}

\begin{document}

\maketitle
\thispagestyle{empty}
\pagestyle{empty}

%%%%%%%%%%%%%%%%%%%%%%%%%%%%%%%%%%%%%%%%%%%%%%%%%%%%%%%%%%%%%%%%%%%%%%%%%%%%%%%%

\begin{abstract}
 Cloud-based battery management systems (BMSs) rely on real-time voltage measurement data to coordinate bi-directional electric vehicle (EV) charging in vehicle-to-grid (V2G) applications.  Unfortunately, an adversary can corrupt the transmitted measurement data, leading to disrupted charging/discharging of EVs.   To ensure reliable voltage data under such sensor attacks, 
    this paper proposes a secure voltage estimation scheme for large-format battery packs based on a self-learning Koopman operator with two-stage error corrections. The first stage compensates for the Koopman approximation error, and the second stage aims to recover the error amassed from the lack of higher-order battery dynamics information in the self-learning feedback. The latter is obtained from two alternative methods: an adaptable heuristic correction that leverages cell-level open-circuit voltage to state-of-charge mapping, and a Gaussian process regression-based correction. We tested our proposed secure estimator using the high-fidelity battery simulation package `PyBaMM-liionpack’, and the results show high accuracy under varying pack topologies, charging settings, battery aging, and attack policies. These findings highlight the scalability and adaptability of our algorithm to diverse battery configurations and operating conditions without requiring significant modifications, excessive data, or sensor redundancy.
\end{abstract}

\section{Introduction}
Large-scale deployment of electric vehicles (EVs) with vehicle-to-grid (V2G) capabilities can balance renewable energy systems and improve grid stability by providing ancillary power services \cite{tavakoli2020impacts}. Effective bi-directional charging of these EVs relies on real-time monitoring and regulation of key battery states such as voltage or state-of-charge \cite{ye2022learning}. Therefore, smart EV charging infrastructures integrate cloud computing and communication technology for optimal V2G charging  \cite{das2024optimal}.
Unfortunately, such cyber-physical battery management systems (BMSs) are susceptible to cyberattacks. For example, corruption of voltage measurement data due to sensor attack
can result in over-charging/discharging of batteries or acute cell-imbalance in battery packs \cite{ghosh2023security}.
These attacks can lead to range anxiety,  unsafe battery conditions, and/or financial loss \cite{acharya2024madeviot}.   Moreover, security vulnerabilities in EV charging stations can be exploited to induce widespread grid disturbances, power outages, over-frequency, and low power factor \cite{rohde2019cyber,morrison2018threats}. This motivates the research on secure battery data for and from BMS to ensure safe and efficient bi-directional V2G charging of EVs as well as reliable grid operations against cyberattacks  \cite{murlidharan2025battery,naseri2023cyber}.  \\ 

To address the security challenges in BMS and EV charging infrastructure,  
\cite{murlidharan2025battery} provided a comprehensive threat assessment to identify the adversarial impact on functionality, safety, and performance of BMS due to several cyberattacks. \cite{murlidharan2025battery} further illustrated how compromised sensing can mislead BMS to make unsafe decisions, resulting in erroneous control actions and poor battery management. In addition to threat assessment,  authors in  \cite{obrien2023detection} proposed an extended Kalman filter-based algorithm to detect false-data-injection (FDI) sensor attacks in battery stacks. Moreover, \cite{dey2020cybersecurity} proposed a model-based dynamic observer to detect cyberattacks on EV charging stations and also showed that such attacks can lead to out-of-service EVs or battery damage via overcharging. \cite{mitikiri2025anomaly} developed a  long short-term memory (LSTM) autoencoder model to detect data spoofing attacks on EV current measurements. Similarly, \cite{rao2023detection} proposed an LSTM model to detect FDI attacks on voltage and current measurements. In contrast, \cite{ghosh2024koopman} proposed a Koopman operator-based algorithm to first detect actuation (charging) attacks and sensor attacks on voltage and temperature measurements, and thereafter, isolate the attack source.

While these detection schemes are vital, it is equally critical to develop mitigation strategies to promptly evade any deleterious impact on EV batteries and widespread grid disturbances \cite{wang2022secure, roy2024input}.  In particular, sensor attacks are designed to deceive cloud-BMS and administrators about the true operating conditions of the system, and in general, are harder to mitigate without sensor redundancy \cite{xu2025distributed}. Failure to mitigate sensor attacks can thus lead to incorrect control actions, poor battery management, reduced battery life, and unpredictable failures such as thermal runaway, fires, or explosions \cite{murlidharan2025battery}. Consequently,  in \cite{tian2022security}, authors proposed an equivalent-circuit model (ECM)-based non-linear estimator for Li-ion batteries to ensure secure state-of-charge (SOC) estimation under compromised sensing.
In \cite{wang2022secure}, authors utilized a time-varying recursive Kalman filter to generate secure SOC estimation under low-intensity deception attacks upon detection.  Similarly, \cite{xiao2025resource} proposed a sensor attack detection and compensation scheme for EV battery systems  and leveraged the discrepancy between set-membership of the state estimation and prediction to estimate the battery SOC under sensor attacks. In \cite{liu2025voltage} authors developed a local outlier factor algorithm to first detect corrupted voltage data (outlier) in power systems, and subsequently,  replaced the corrupted data with estimated voltage to compensate for the attacks. Moreover, in \cite{xu2025distributed}, authors first developed an optimized partitioning and sensor placement strategy for large-scale power systems and proposed an  improved unscented Kalman filter-based distributed secure estimation framework. \cite{dong2025secure} adopted an impulsive-driven observer to estimate SOC that acquires measurement data at discrete instants rather than using a continuous data stream to achieve improved estimation accuracy under compromised sensing. Such a discrete measurement feedback strategy reduces the susceptibility to attacks and also minimizes the cumulative impact of compromised data on estimation.

Though these studies address some of the crucial needs for resilient sensing in batteries under sensor attacks, they also reveal several key research gaps. These works adopt model-based approaches and thus require reliable identification of model parameters for different batteries to ensure accurate estimation \cite{trevizan2022cyberphysical}. Additionally, model-based algorithms have higher computational complexities and thus scale poorly to larger packs \cite{ghosh2025detection}.
In this context, our previous paper \cite{ghosh2025secure} presents a proof of concept of a self-learning feedback mechanism for Koopman operator (KO) based secure estimation without measurement feedback in battery systems. 
Using this self-learning KO-based secure estimator, we obtained accurate long-term voltage estimation under compromised sensing for a single-cell Li-ion battery. This preliminary work explored the plausibility of this framework for Li-ion batteries during charging by leveraging the data generated from an equivalent circuit model (ECM). However, there is a research gap to assess the scalability of this proposed self-learning KO strategy to large-format battery packs using high-fidelity data at a realistic sampling frequency.
Thus, our preliminary work \cite{ghosh2025secure} motivates the research scope for further explorations to substantiate the applicability of our secure estimation algorithm in real-world scenarios.
Our contribution addresses these gaps in the following way: %\textcolor{red}{recheck}
\begin{enumerate}
    \item We proposed a two-stage error compensated self-learning KO-based secure voltage estimator for large-format battery packs under sensor cyberattacks. 
    This model-free secure estimator adopts a small-data online learning approach that requires only limited model information and does not rely on sensor redundancy.
    \item We propose a scalable battery characteristic-guided heuristic error compensation strategy that leverages the cell-level OCV-SOC map for pack-level estimation. % without any added computational burden. 
    Alternatively, we propose a GPR-based data-driven error correction that bypasses the empirical derivation of the heuristic function using battery characteristic information.
    \item We validate the applicability of the proposed secure estimator for battery packs with varied pack topologies and degradation levels under both charging and discharging conditions. We use the rate-limited voltage measurements and current input data from the high-fidelity simulation package `PyBaMM-liionpack' \cite{tranter2022liionpack} for our experiments.
\end{enumerate}

The rest of the paper is organized as follows. Section \ref{probForm} presents the problem framework of this paper. We introduce our proposed self-learning Koopman operator-based secure estimation algorithm along with two-staged error compensation strategies in Section~\ref{algo}. Our simulation results are presented in Section~\ref{sim}. Finally, Section \ref{conclu} concludes our paper.

%%%%%%%%%%%%%%%%%%%%%%%%%%%%%%%%%%%%%%%%%%%%%%%%%%%%%%%%%%%%%%%%%%%%%%%%%%%%%

\section{Problem Formulation} \label{probForm}
This paper addresses the problem of secure voltage estimation during bi-directional V2G charging of EVs  under compromised sensing. In this section, we first describe our framework and  battery dynamics under sensor attacks.
Next, we present a brief overview of the Koopman linear approximation of battery dynamics and the event-triggering mechanism that activates the secure estimator under voltage sensor attacks.

\subsection{Bi-directional V2G charging  under compromised sensing} 
\begin{figure}[h]
    \centering
    \includegraphics[width = 0.7\linewidth]{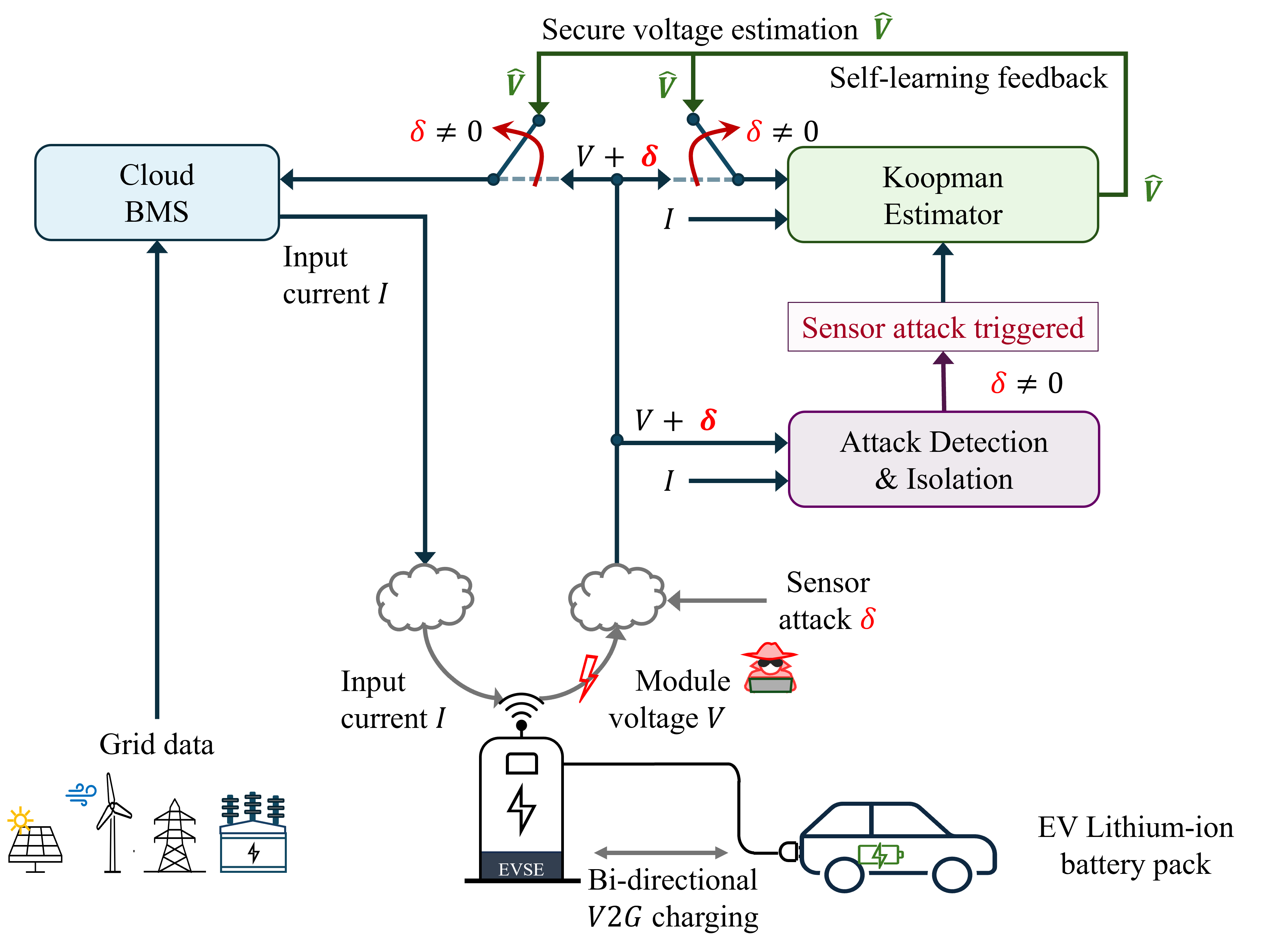}
    \caption{Figure shows  the secure voltage estimation generation under sensor attack.}
    \label{fig:overallB}
\end{figure}

In this paper, we consider that a cloud-based BMS monitors voltage measurements to provide an appropriate current actuation command to the local-BMS to support bi-directional V2G charging.  The voltage measurements from the local-BMS to the cloud-BMS can be corrupted by sensor cyberattacks, while the actuation commands from the cloud-BMS to the local-BMS can be corrupted with actuation attacks during transmission \cite{johnson2022cybersecurity}. Using \cite{ghosh2025detection}, the presence of sensor attacks can be detected and isolated. Upon detection of sensor attack $\delta$ on battery voltage $V$, in this paper, we aim to estimate voltage $V$ as $\hat{V}$. Fig.~\ref{fig:overallB} shows the overview of our objective of secure voltage estimation in the presence of a sensor attack. Under voltage sensor attack $\delta$, battery  dynamics can be written as: 
\begin{align}\label{vt_dynamics}
      x(k+1) = f(x(k),{I}(k)); \,\, V(k) = g(x(k)) + \delta (k).
\end{align}
%\noindent
Here, $x(k) \in  \mathbb{R}^d$ represents the internal battery states at $k^{th}$ instant that may include lithium-ion concentrations, solid and liquid phase potentials, open circuit potential, etc. \cite{sulzer2021python}. $V(k) \in \mathbb{R}^m$ is the vector of module voltage measurements at the $k^{th}$ instant as $V (k) = \begin{bmatrix}
    V^1 (k) & \cdots & V^m (k)
\end{bmatrix}^T$, where $m$ is the total number of modules in the pack.  $\delta (k)$ is the sensor cyberattack that corrupts the voltage measurements at the $k^{th}$ instant.
 ${I} \in \mathbb{R}$ is the input current signal such that during discharging ${I} >0$ and during charging ${I} < 0$.   The non-linear function $f: \mathbb{R}^d \times \mathbb{R} \rightarrow \mathbb{R}^d$ captures the internal dynamics of the battery, and $g: \mathbb{R}^d  \rightarrow \mathbb{R}^m$ defines the non-linear measurement function.  

 Next, we outline the implementation of the Koopman operator (KO) based learning of battery dynamics \eqref{vt_dynamics}.

\subsection{Koopman-operator for battery module voltage prediction} \label{Koopman model}
\begin{figure}[h]
    \centering
    \includegraphics[width = 0.9\linewidth]{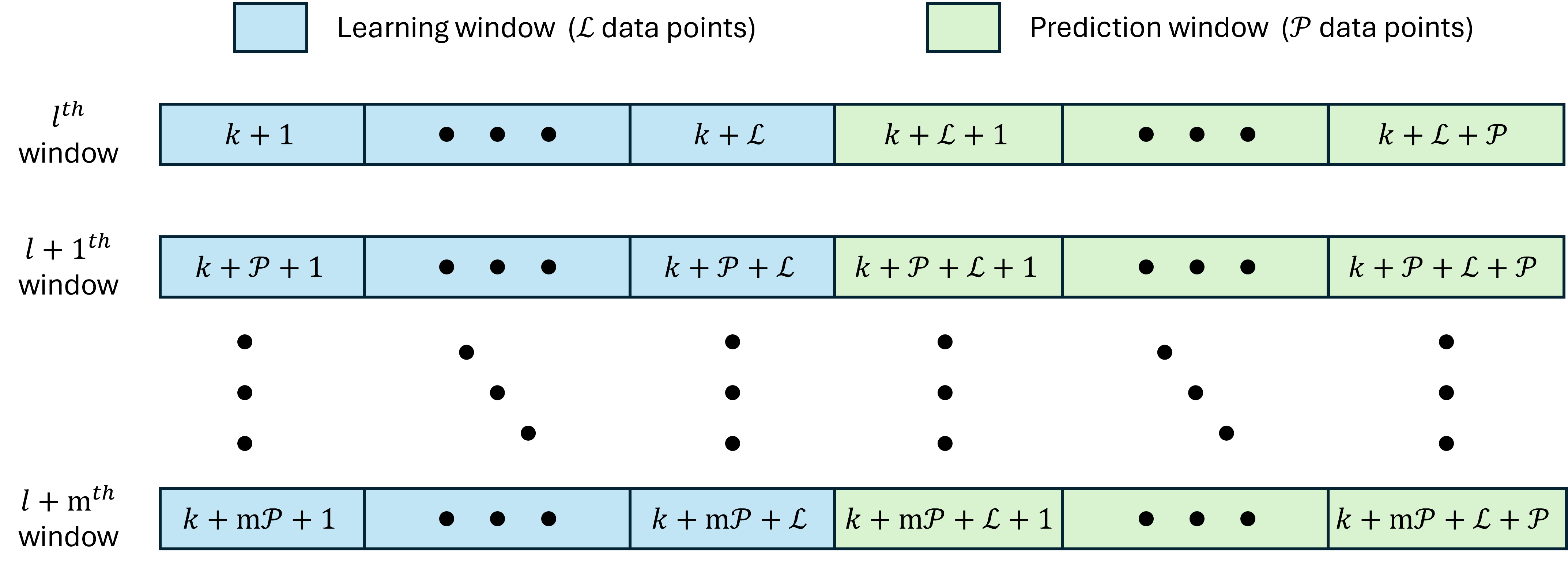}
    \caption{Diagram shows the sliding window mechanism for learning of Koopman parameters and generating module voltage prediction ${V_p}$ as \eqref{g in phi}.}
    \label{fig:sliding}
\end{figure}

%\noindent
We utilize KO theory combined to obtain a transformation $x \mapsto z$ such that the \textit{nonlinear} battery dynamics of states $x$ can be embedded in a \textit{ linear} state dynamics in $z$ \cite{korda2018linear}. 
To define the KO for the battery system \eqref{vt_dynamics}, let us 
first define a right shift operator $\mathbf{S}: \mathbb{R} \rightarrow \mathbb{R}$ such that $\mathbf{S} \mathbf{I} (k) = \mathbf{I}(k+1), \forall k$, where $\mathbf{I} (k)$ is the $k^{th}$ element of the sequence $\{{I}(k)\}_{k=0}^{\infty}$. 
Now, we consider an infinite-dimensional Hilbert space  $\mathcal{H}$ such that $\psi\in \mathcal{H}$ is a complex-valued observable function and $\psi: \mathbb{R}^d \times \mathbb{R} \rightarrow \mathbb{C}$. Then for any observable  $\psi$, state $x$,  and  input ${I}$, a KO  $\mathcal{K} : \mathcal{H} \rightarrow \mathcal{H}$  is defined   as
\begin{align}
\centering
    (\mathcal{K} \psi)(x(k), \mathbf{I}(k)) & = \psi \big( f(x(k),{I}(k)), \mathbf{S} \mathbf{I} (k)\big)\nonumber\\
    &= \psi \big(x(k+1), \mathbf{I}(k+1)\big) \coloneqq \psi_{k+1}. \label{Kt_psi}
\end{align}
\eqref{Kt_psi} implies that learning $\mathcal{K}$ allows us to predict the observable (measurement) $\hat{\psi}_{k+1}$ from $\psi_k$.
It can be easily chosen that $\mathcal{K}$ in \eqref{Kt_psi} is a linear operator. Since $\mathcal{H}$ is the Hilbert space of observables, $\mathcal{H} =\text{span} \{\phi_i \}_{i = 1}^\infty$, where $\phi_i: \mathbb{R}^d \times \mathbb{R} \rightarrow \mathbb{C}$ are the eigen-observables. Therefore, KO $\mathcal{K}$ evolves these eigenfunctions $\phi_i$ linearly in time with the corresponding Koopman eigenvalues, $\lambda_i \in \mathbb{C}$ as \cite{brunton2019notes}:
\begin{align}
 \mathcal{K} \phi \big(x(k), \mathbf{I}(k)\big) = \lambda  \phi \big(x(k), \mathbf{I}(k)\big). \label{KO for Phi}
\end{align}
Then  ${{\psi}}  \in \mathcal{H}$  can be expressed as  $ \psi_{k+1} = \sum_{i=1}^{\infty} \lambda_i \phi_i \big (x(k), \mathbf{I} (k) \big) v_i^{\psi}$, where the Koopman Modes (KM) $v_i^{{\psi}} \in \mathbb{C}$ are the coefficients of the projection of ${{\psi}}(x)$ onto the span$\{\phi_i\}_{i = 1}^\infty$  \cite{mezic2005spectral}.  For practical implementation, if we can find a finite set of $\{\phi_i\}_{i = 1}^n$ to span a Koopman invariant subspace such that if $\psi$ lies in this subspace, then an $n^{th}$ dimensional approximation can be achieved by 
\begin{align}\label{psi in phi}
   \psi_{k+1} \approx \sum \limits_{i=1}^{n} \lambda_i \phi_i \big (x(k), \mathbf{I} (k) \big) v_i^{\psi}.
\end{align}
Furthermore, if the voltage measurement function $g$ in \eqref{vt_dynamics} lies in the Koopman invariant subspace spanned by $\{\phi_i\}_{i = 1}^n$, a good approximation can be achieved by 
\begin{align}\label{g in phi}
   g\big(x(k+1) \big) \approx {V_p}(k+1) = \sum \limits_{i=1}^{n} \lambda_i \phi_i \big (x(k), \mathbf{I} (k) \big) v_i^{g}.
\end{align} 
Here, ${V_p}(k+1)$ is the  module voltage prediction at the $k+1$-th time instant. We adopt delay embedding along with dynamic mode decomposition (DMD) techniques to obtain such a finite-dimensional Koopman approximation of battery dynamics \eqref{vt_dynamics}, where Koopman parameters $\lambda_i$, $\phi_i$, and $v^g_i$ are learned using module voltage data $V$ and input current signal $I$ over a sliding learning window  of length $\mathcal{L}$. These parameters are then used to generate voltage prediction ${V_p}$ for the next $\mathcal{P}$ time instants using \eqref{g in phi}. The learning and prediction windows move forward $\mathcal{P}$ instants after every prediction cycle, and Fig~\ref{fig:sliding} illustrates the sliding mechanism of the windows.

Thus, in the nominal scenario, we use the module voltage data $V$ and the input current signal $I$ to learn the Koopman linear battery model.
However, in the presence of a sensor attack $\delta \neq 0$, the voltage measurement $V$ is not admissible. Therefore, we propose an event-triggered secure voltage estimation strategy that adopts a self-learning Koopman approach. The secure estimator uses voltage prediction $V_p$ as self-feedback for KO learning to avoid the adverse impact of corrupted voltage measurement. Next, we briefly describe the self-learning triggering mechanism.

\subsection{Identification of sensor attacks}
%\noindent 
We utilize a KO-based diagnostic algorithm \cite{ghosh2025detection} to reliably identify the presence of sensor attacks and activate the secure estimator. The diagnostic algorithm utilizes the module voltage data $V$ and the input current signal $I$  to learn the Koopman parameters over a learning window and generates voltage predictions $V_p$ as \eqref{g in phi} for the following prediction window.
The algorithm monitors the detection residual $RD (k) = V(k)- V_p (k)$, i.e., the difference between the true and predicted voltage data to capture the presence of an attack, and an attack decision is made when the residual $RD$ crosses a predefined threshold. Next, the algorithm utilizes the distinct signatures of actuation and sensor attacks, respectively, on the Koopman eigenfunctions $\phi_i$ and modes $v_i^g$ for isolation. In particular, the algorithm monitors the deviation in KMs $v_i^g$ due to the presence of an attack as the isolation residual $RI$, and in the presence of the sensor attack $\delta \neq 0$, $RI$ crosses a predefined threshold to raise a sensor attack flag.  Once  a sensor attack is identified, the proposed secure estimator is triggered.

 \section{Secure Module Voltage Estimation Algorithm} \label{algo}
In this section, we present the proposed self-learning  KO-based secure estimation algorithm. Under nominal scenario ($\delta =0$), we use available voltage data $V$ and current data $I$ to effectively learn the Koopman parameters and generate module voltage predictions online. However, when the diagnostic algorithm identifies a sensor attack and triggers the secure estimation algorithm ($\delta \neq 0$), corrupted module voltage measurements are excluded from updating the Koopman operator. Instead, the secure estimator starts to utilize the Koopman prediction $V_p$ as feedback to itself, i.e., it initiates self-learning feedback. 
Thus, during secure estimation in the presence of sensor attack $\delta \neq 0$, 
the Koopman voltage predictions $V_p$ from the $l$-th window are utilized as self-feedback to learn the Koopman parameters $\lambda_i$, $\phi_i$, and $v^g_i$ in the subsequent windows. For both nominal and secure estimation operations, we utilize the current input data $I$ for learning and prediction. 

Nevertheless, the implementation of such self-learning KO introduces two drawbacks.  Firstly, since the secure estimator receives only Koopman voltage predictions $V_p$ for self-learning feedback, it accumulates prediction errors from the Koopman linear approximation. Next, the secure estimator will cease to capture any higher-order changes in battery dynamics due to the lack of new system information in self-learning feedback. This leads to a further accumulation of prediction errors in self-learning feedback. To address these challenges, we propose two-stage error correction for the self-feedback to reduce error accumulation from the Koopman approximation (stage~I) and the higher-order battery dynamics (stage~II).

 \subsection{Stage~I: Koopman approximation error correction} 
 %\noindent
This first stage of correction estimates the potential error in voltage predictions $V_p$ to compensate for inaccuracies from finite-dimensional Koopman approximation in \eqref{g in phi}. 
Hence, we deploy parallel KO learning to estimate the prediction error  $E_p (k) = V_{nom} (k) - V_p(k)$, where $V_{nom} (k)$ is the true voltage measurement at the $k^{th}$ instant. We adopt a similar delay embedded DMD strategy to learn the error dynamics from the error $E_p$ data  and the input $I$ data. 
Then we utilize the approximated KO for the error dynamics to generate $E_1$, where $E_1$ is the estimation of the prediction error $E_p$.

True measurement $V_{nom}$ is only available during nominal operation ($\delta = 0$). Therefore, once the sensor attack is detected and secure estimation is triggered, this parallel Koopman for error dynamics is similarly self-learned using the estimated error $E_1$.
The prediction $E_1$ from the Koopman error dynamics is then added to the predicted voltage $V_p$ to obtain:
\begin{align}
    (\text{Stage~I correction})\,\quad \overline{V}_p(k) = V_p(k)  + E_1(k). \label{stage I corr}
\end{align}
%\noindent
This stage~I corrected voltage estimation $\overline{V}_p$ is subsequently used as self-learning feedback for the Koopman voltage predictor, i.e., KO is approximated using voltage estimation $\overline{V}_p$ data and input $I$ data over the learning window. The Koopman approximation error compensation $E_1$ while largely improves the self-learned voltage estimation $\overline{V}_p$, it still fails to incorporate any information that can capture higher-order battery dynamics. This leads to a large under- or over-estimation error in the stage~I corrected voltage estimation $\overline{V}_p$. Thus, we add stage~II compensation $E_2$ to obtain the final error-compensated secure voltage estimation $\widehat{V}$. 
 In subsequent sections, we introduce two alternative methods: one heuristic strategy and one data-driven strategy to compute stage~II error compensation $E_2$ from stage~I error compensation $E_1$.

 \subsection{Heuristic Stage~II: Higher-order dynamics correction with OCV-SOC mapping} 
%\noindent
This method  computes the error compensation $E_2$
from the current input and the OCV-SOC mapping using a heuristic approach based on empirical battery knowledge. 
 From physics-based battery models such as equivalent circuit or electrochemical models, we know that the battery terminal voltage has a strong relation with the battery OCV through the current or the SOC. 
 Therefore,  we consider that higher-order dynamics in true voltage $V_{nom}$ can be captured through the higher-order characteristic changes in the OCV-SOC mapping.
 For this method, we utilize the nominal battery capacity $Q$, OCV vs SOC map, and initial SOC, which are more readily accessible without requiring sensor redundancy.

 %\noindent
 Our numerical analysis on Koopman operator estimation shows that the error accumulation in the stage~I compensated estimate $\overline{V}_p$ varies across different SOC regions. Specifically, this finding also matches the characteristic changes in the OCV-SOC map. We further observe that the current flow direction during the charging vs discharging of the battery also impacts the error accumulation. Based on our observation,
 we first adopt the Coulomb counting method to calculate the SOC from the current ${I}$,  capacity $Q$, and initial $SOC(0)$ data as $SOC ({k+1}) = SOC(k) + (\Delta t\,   {I})/Q$, where $\Delta t$ is the sampling time. Next, the calculated SOC  and the OCV-SOC map are utilized to track changes in OCV, and in turn, estimate the drift in predicted voltage $\overline{V}_p$  due to higher-order dynamics in true voltage $V_{nom}$. Thus, we propose the following algebraic formula to obtain the OCV-SOC map-dependent heuristic correction  as follows:
 \begin{align} \label{total_compen}
    E_2 (k) &=  h (SOC (k), sgn({I(k)}))  E_1 (k) \nonumber \\
    & \quad \,\,\,\, + \left[1-SOC (k)\right] \Delta OCV (k).
 \end{align}
Here, the function $h (SOC, sgn({I}))$ exhibits a piecewise constant mapping to SOC that varies with the charging direction $sgn({I})$. The sign function $sgn({I})$ returns 1 for discharging and -1 for charging. Here, $\Delta OCV (k) = OCV(k) - OCV(k-1) $ is the change in OCV with battery operation. 
 To learn this $h (SOC, sgn({I}))$ function, we first use the $\frac{d^2\,(OCV)}{d\,SOC^2 }= 0$ and $\frac{d^3\,(OCV)}{d\,SOC^3 }= 0$ points to divide the SOC range into $N_j$ regions. While the OCV slope, i.e., the first-order derivative $\frac{d\,(OCV)}{d\,SOC}$ remains strictly non-negative (during charging) or non-positive (during discharging), $\frac{d^2\,(OCV)}{d\,SOC^2 }= 0$ and $\frac{d^3\,(OCV)}{d\,SOC^3 }= 0$ points, respectively, reflect the changes in the OCV slope and curvature. These second- and third- order critical points capture the changes in OCV with SOC, and we obtain the SOC regions based on these points to incorporate such higher-order characteristic shifts into our stage~II correction.  For each region, we heuristically deduce the $h (SOC, sgn({I}))$ mapping such that $E_2$ can capture the higher-order variability in true voltage $V_{nom}$.
This error compensation $E_2$ is added to the Koopman voltage prediction $V_p$ to obtain:
\begin{align}
    (\text{Heuristic Stage~II correction})\,\,\,\, \widehat{V} (k) = V_p (k) + E_2 (k). \label{emp corr}
\end{align}
\begin{figure}[b]
    \centering
     \includegraphics[width=0.67\linewidth]{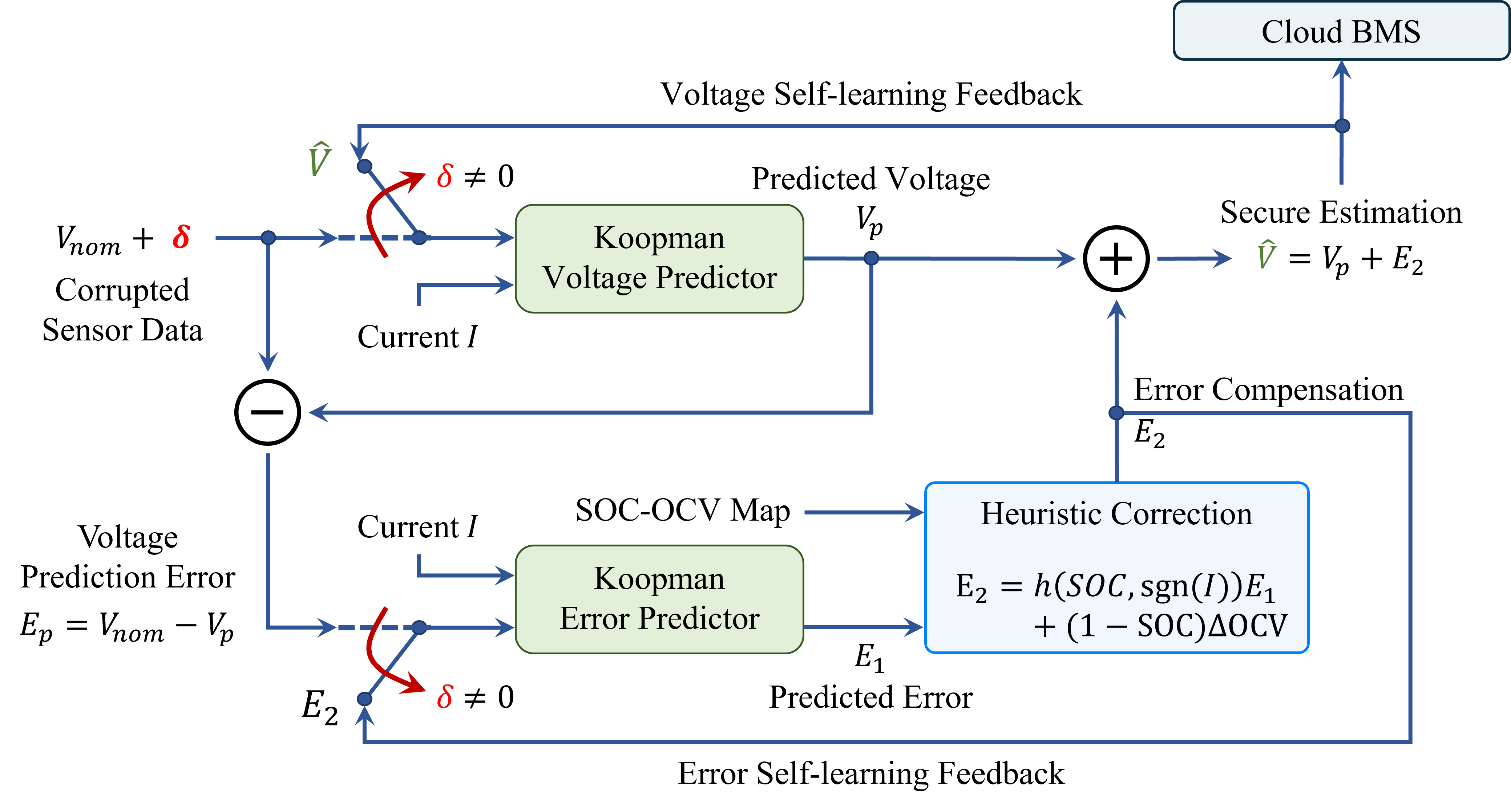}
    \caption{Diagram shows the mechanism of secure voltage estimation with heuristic stage~II correction. }
    \label{fig:heu}
\end{figure}
The secure voltage estimate $\widehat{V}$ is sent to the cloud-BMS as well as to the Koopman operator as self-learning feedback. 
Similarly, to compensate for the inaccuracies due to the lack of new information in the self-learned Koopman-based error estimate $E_1$, 
error compensation $E_2$ is used as self-learning feedback to the Koopman error predictor.
Fig.~\ref{fig:heu} illustrates the generation of heuristic stage~II error compensation $E_2$ using \eqref{total_compen} and the secure voltage estimate $\widehat{V}$, which are used as self-learning feedback to the Koopman error predictor and Koopman voltage predictor, respectively.

 \subsection{Data-driven Stage~II: Higher-order dynamics correction with GPR} 
For this data-driven method, we adopt Gaussian process regression (GPR) to directly approximate the comprehensive error compensation $E_2$. GPR is a non-parametric supervised regression technique that learns nonlinear input–output relationships by modeling the target function as a Gaussian process. 
It assumes that the probability distribution for any inputs $\theta$ over a function $G(\theta)$ exhibits a Gaussian distribution and can be described as ${G}(\theta) \sim GP(M(\theta), \kappa_G (\theta_i, \theta_j))$. Here $M(\theta)$ is the mean function, and $\kappa_G$ is a pairwise kernel function that measures the similarity between two input points $\theta_i$ and $\theta_j$.
Thus, GPR provides a probabilistic prediction that includes both the predicted mean and the associated uncertainty for each prediction \cite{liu2019modified}
% attempts to find the underlying structure of the data points while providing the predictions as probability distributions \cite{liu2019modified}. In addition,
This makes GPR particularly effective in cases where the relationship between the input and output variables is unknown or complex.   Hence, we choose GPR to approximate the error compensation $E_2$.

\begin{figure}[h!]
    \centering
    \includegraphics[width=0.67\linewidth]{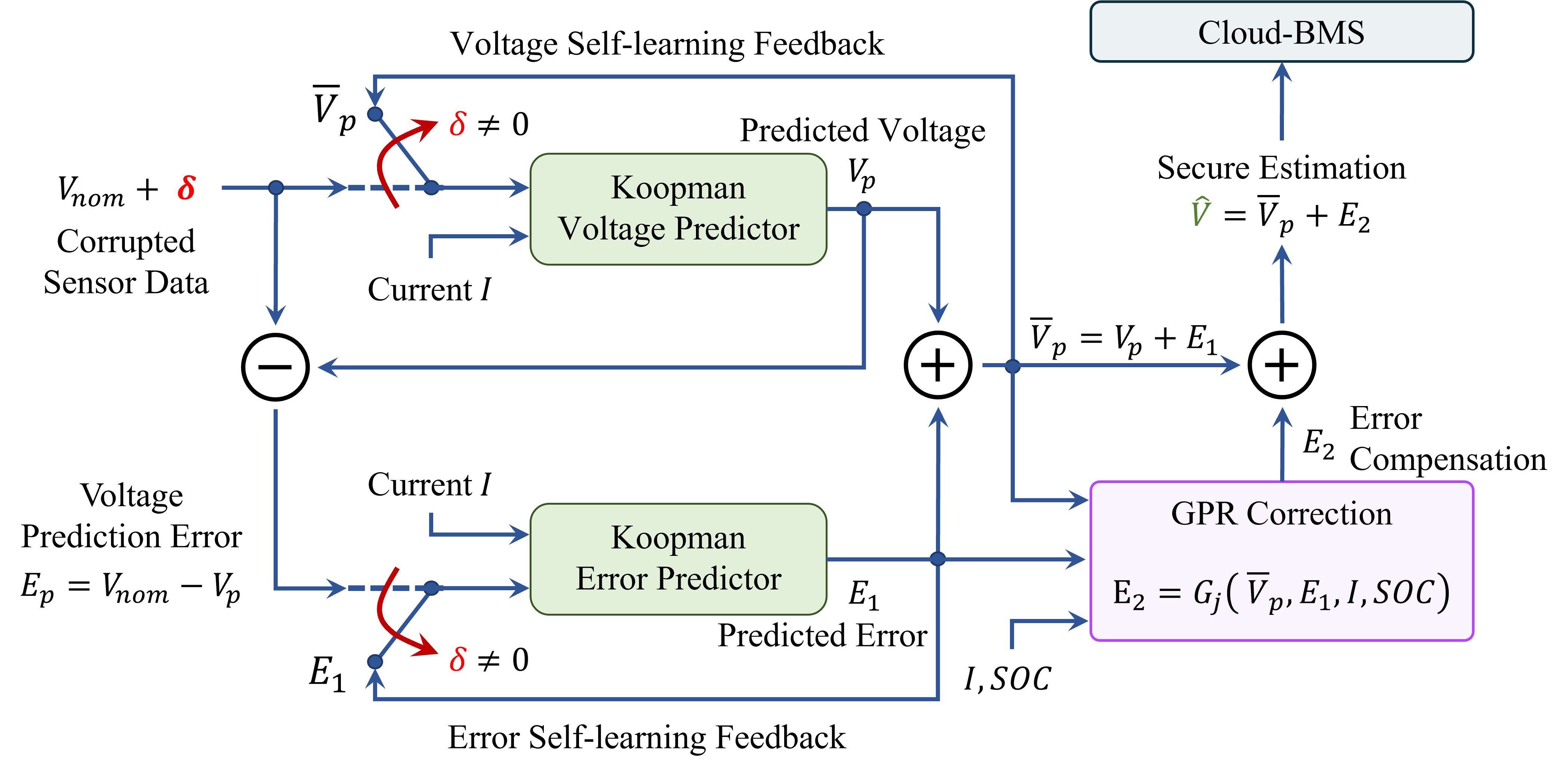}
    \caption{Diagram showing the mechanism of secure voltage estimation with GPR stage~II correction.} 
    \label{fig:gpr_m}
\end{figure}

%\noindent
We utilize the error in the stage~I compensated voltage estimation $\overline{V}_p$ to approximate the compensation $E_2$ with our GPR model. Hence, the target data for our GPR model is $E_2 (k) = V_{nom} (k) - \overline{V}_p (k)$, 
corresponding to four input features: self-learned approximation error compensation $E_1$,  stage~I corrected estimation $\overline{V}_p$, current ${I}$, and calculated SOC, i.e., $\theta (k) = \begin{bmatrix}
   E_1^T (k)  &  \overline{V}_p^T (k) & {I}(k) & SOC (k)
\end{bmatrix}^T$. We train the GPR model to learn the latent mapping from the input $\theta$ to the compensation $E_2$ as $E_2 (k) = G(\theta(k))+\eta (k), \forall k$, where $\eta \sim \mathcal{N}(0,\sigma_\eta^2)$ is zero-mean Gaussian noise. We adopt the squared exponential kernel function \eqref{SE kernel} and a constant basis mean function as $M(\theta) = \beta$ to learn our GPR model $\mathcal{G}$. 
\begin{align}
    \kappa_G (\theta_i, \theta_j) = \sigma_G^2 \,\,\text{exp} \left(-\frac{(\theta_i - \theta_j)^2}{2 \,L^2} \right). \label{SE kernel}
\end{align}
These 4 hyperparameters $\beta, \sigma_G, L, \sigma_\eta$ are optimized with the Quasi-Newton method. % to learn the GPR model $\mathcal{G}$. 
Furthermore, since our empirical findings highlight the varied error accumulation across different SOC regions due to higher-order battery dynamics, we similarly consider $N_j$ regions in the SOC range for this method. 
Then, a total of $N_j$ GPR models are trained to obtain the most accurate voltage estimates, and we utilize  GPR model  $\mathcal{G}_j$ for the $j$-th SOC region to estimate the secure voltage error compensation $E_2$ as
\begin{align}
    E_2 (k) = \mathcal{G}_j (\theta (k)).  \label{gpr models}
\end{align}
This error compensation $E_2$ is then used  to generate the secure estimation as 
\begin{align}
    (\text{GPR Stage~II correction})\,\quad \widehat{V}(k) =  \overline{V}_p (k)  + E_2 (k). \label{GPR corr}
\end{align}
This GPR Stage~II corrected secure estimation $\widehat{V}$ is  sent to cloud-BMS to ensure safe battery operation; however, 
the Koopman voltage predictor continues to run with stage~I correction using ${\overline{V}_p}$ data as self-learning feedback. Fig.~\ref{fig:gpr_m} shows the generation of GPR stage~II error compensation $E_2$ using \eqref{gpr models} and the secure voltage estimate $\widehat{V}$, along with $E_1$ and ${\overline{V}_p}$ that are used as self-learning feedback to the Koopman error predictor and Koopman voltage predictor, respectively.

\section{Simulation Results} \label{sim}
\subsection{Data generation with Pybamm-liionpack}\label{dataGen}
%\noindent
We have utilized the `liionpack' extension of the open-source battery simulator PyBaMM (Python Battery Mathematical Modeling) to generate our V2G charging/discharging scenarios \cite{tranter2022liionpack}. The PyBaMM-liionpack framework has emerged as a reliable and powerful tool tailored for efficient high-fidelity battery simulations  \cite{ghosh2025transfer}. For our case studies, we utilize the Single-Particle Model (SPM) \cite{atlung1979dynamic} in liionpack to conduct our battery experiments for the commercial cylindrical battery cell LGM50 \cite{chen2020development}. The nominal discharge capacity of these cells is $5 \, Ah$  between the lower and upper cut-off voltages of $2.5 \, V$ and $4.2 \, V$, respectively.  We have evaluated our secure estimator algorithm on two large-format battery packs aligned with standard EV battery specifications \cite{sequino2021modeling}.
    \begin{enumerate}
    \item We consider a battery pack of  900 LGM50 cells, and the cells are configured in 3 parallel modules with 5p60s format, i.\,e.  60 series branches of 5 parallel cells.  The maximum voltage of the pack is 250 $V$, and the nominal discharge capacity is 75 $Ah$ with 16$kWh$  energy capacity.
    \item The second battery pack contains  2000 LGM50 cells arranged in 4 parallel modules with 5p100s format.  The maximum pack voltage is 420 $V$ with nominal discharge and energy capacities of 100 $Ah$ and 35$kWh$, respectively.
    \end{enumerate}
    For all of our experiments, we adopt the constant-current charging and discharging policy with 1C charging rate for all packs. To generate battery aging data, we repeatedly charge and discharge the battery packs between 0.3 and 0.7 SOC for 100 cycles, and we consider resting the battery for 15 minutes after each charging or discharging phase. Moreover, we sample our module voltage measurement and input current data at a $1Hz$ rate, i.e., one sample per second.

\subsection{OCV-SOC based SOC region partition}
\begin{figure}[h!]
      \centering
      \includegraphics[width=0.7\linewidth]{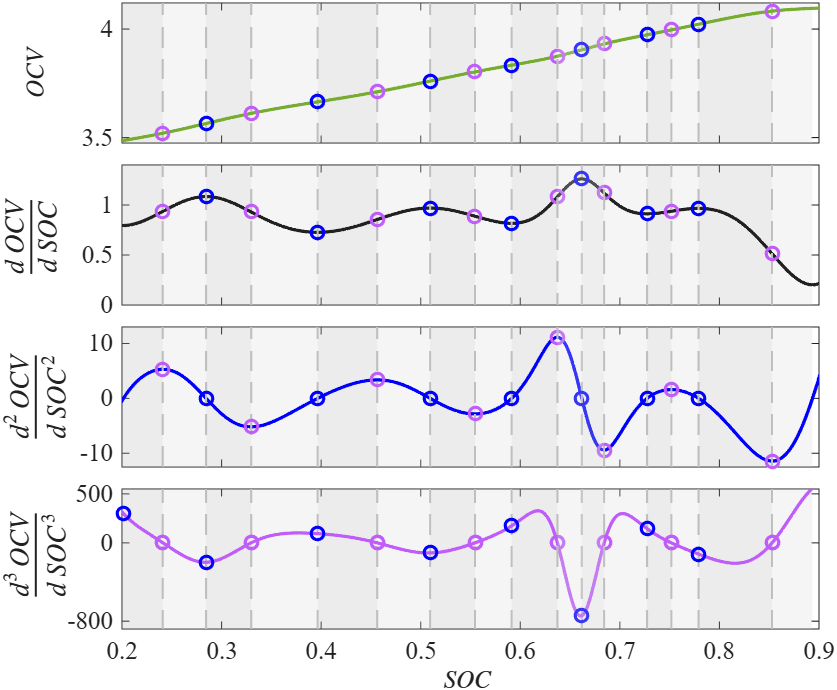}
      \caption{For the LGM50 Lithium-ion battery, the figure exhibits OCV-SOC  (top), $\frac{d\,OCV}{d\,SOC}$ (second), $\frac{d^2\,OCV}{d\,SOC^2}$ (third), and $\frac{d^3\,OCV}{d\,SOC^3}$ (bottom) graphs. }
      \label{fig:soc_ocv}
  \end{figure}

%\noindent
Determining the appropriate SOC regions is crucial to ensure reliable stage~II correction using both heuristic and data-driven methods for our proposed secure estimator. First, to obtain the OCV-SOC map for the LGM50 battery cell, we first slowly charge the battery cell with $0.05 A$ ( $(\frac{1}{100})^{th}$ C-rate) and measure the terminal voltage once per second to capture the OCV at the corresponding SOC \cite{lee2008state}. Here, the SOC is calculated using the Coulomb counting method.  We then utilize the $\frac{d^2\,OCV}{d\,SOC^2 }= 0$ and $\frac{d^3\,OCV}{d\,SOC^3 }= 0$ points 
% to identify the changes in the slope and curvature of the OCV-SOC map, respectively. Next, these points are used t
to partition the SOC range into 16 regions, while capturing the higher-order characteristic changes in the OCV-SOC map. Fig.~\ref{fig:soc_ocv} presents the OCV-SOC curve (top), $\frac{d\,OCV}{d\,SOC}$ graph (second), $\frac{d^2\,OCV}{d\,SOC^2}$ graph (third), and $\frac{d^3\,OCV}{d\,SOC^3}$ graph (bottom) for this battery. In addition, the dotted line in each plot of Fig.~\ref{fig:soc_ocv} marks the 16 SOC regions, also highlighted by alternating shaded backgrounds. We have also indicated the $\frac{d^2\,OCV}{d\,SOC^2 }= 0$ corresponding boundary points with blue circles and the $\frac{d^3\,OCV}{d\,SOC^3 }= 0$ corresponding points with purple circles. Next, we describe the region-specific learning of the heuristic and GPR stage~II correction methods to incorporate these OCV characteristic changes into our proposed secure estimator.
% 
   % \subsection{}
% 
  \subsection{Region-specific learning of heuristic stage~II correction:}
   We use a brute-force strategy to derive the piecewise-constant values of $h (SOC, sgn({I}))$ for each of the 16 SOC regions of the LGM50 battery cell.  Table~\ref{tab:ls} lists these values of $h (SOC, sgn({I}))$ for each SOC region $j$ with the interval $\mathcal{I}_{s,j}$, where $h (SOC, 1)$ and $h (SOC, -1)$ correspond to discharging and charging phase, respectively.  Most importantly, the intrinsic OCV-SOC characteristic exhibits only marginal deviation from a battery cell to a series-connected module, where the module OCV magnitude at a given SOC  scales with cell count \cite{yu2023ocv}.
   Hence, we utilize the same $h (SOC, sgn({I}))$ values learned from cell-level analysis, while only scaling $\Delta OCV$ part in \eqref{total_compen} for all battery packs. We multiply the cell-OCV by the total number of series-connected cells in a module to scale the cell-level OCV-SOC map to that battery pack. 
     \renewcommand{\arraystretch}{1.3}
\begin{table}[h!]
      \centering  
      \caption{Piecewise-constant values of $h (SOC, sgn(I_c))$ required for heuristic correction}
      \begin{tabular}{|c|l|c|c|}
      \hline
         { SOC region } &  \multirow{2}{*}{Interval $ \mathcal{I}_{s,j}$} & $h (SOC, 1)$  & $h (SOC, -1)$ \\
        $j $ & & (Discharging) & (Charging) \\
          \hline
          1 & $ [0, 0.241)$ & 0.989 & 0.960 \\
          \hline
          2 & $ [0.241, 0.284)$ & 0.853 & 0.948 \\
          \hline
          3 & $ [0.284, 0.330)$ & 0.952 & 0.952 \\
          \hline
          4 & $ [0.330, 0.397)$ & 0.989 & 0.968 \\
          \hline
          5 & $ [0.397, 0.456)$ & 0.955 & 0.955 \\
          \hline
          6& $ [0.456, 0.510)$ & 0.946 & 0.945 \\
          \hline
          7 & $ [0.510, 0.555)$ & 0.883 & 0.960\\
          \hline
          8 & $ [0.555, 0.591)$ & 0.990 & 0.922 \\
          \hline
          9 & $ [0.591, 0.6373)$ & 0.999 & 0.990\\
          \hline
          10 & $ [0.6373, 0.662)$ & 0.999 & 0.990\\
          \hline
          11 & $ [0.662, 0.684)$ & 0.963 & 0.978 \\
          \hline
          12 & $ [0.684, 0.727)$ & 0.963 & 0.978 \\
          \hline
          13 & $ [0.727, 0.752)$ & 0.911 & 0.920\\
          \hline
          14 & $ [0.752, 0.779)$ & 0.960 & 0.860\\
          \hline
          15 & $ [0.779, 0..853)$ & 0.967 & 0.880 \\
          \hline
          16 & $ [0.853, 1)$ & 0.945 & 0.920 \\
          \hline
      \end{tabular}
      \label{tab:ls}
      % \vspace{-4mm}
  \end{table}

  %\noindent
  % Next, 

    \subsection{Region-specific learning of GPR stage~II correction:}
    We learn 16 GPR models corresponding to 16 SOC regions to ensure reliable and accurate secure voltage estimation for a battery module. We use the function `fitrgp' to train our GPR models in MATLAB R2024b, while considering the squared-exponential kernel \eqref{SE kernel}.  
  The GPR models exhibit poor generalization to different pack configurations.
   Therefore, we separately train 16 region-specific GPR models for each battery and obtain a total of 32 GPR models, i.e., two sets of 16 region-specific models corresponding to two battery packs. 

  \subsection{Estimation performance assessment}
\begin{figure}[h]
    \centering
    \includegraphics[width=0.68\linewidth]{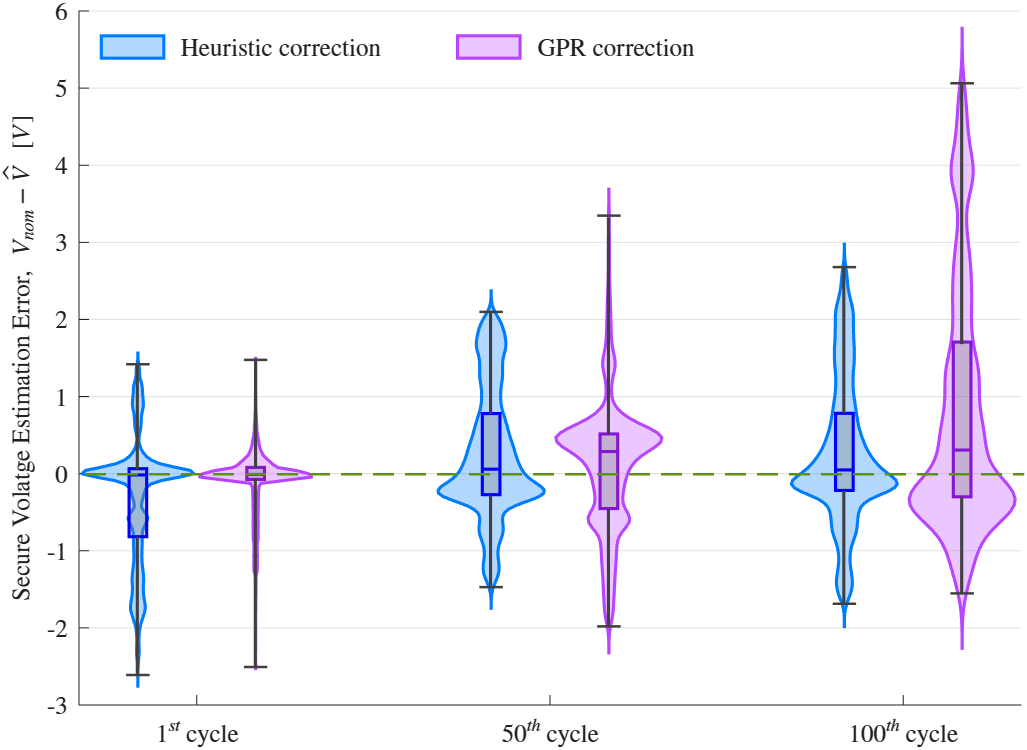}
    \caption{ Plot compares the distributions of voltage estimation error for heuristic and GPR correction at three degradation levels of the battery.  }
    \label{fig:violin}
\end{figure}

  %\noindent
  Reliable, accurate, and real-time voltage estimation with improved adaptability is essential to ensure safe and optimum control of EV charging operations. Thus, we conduct an extensive Monte Carlo simulation of 1090 test runs where we evaluate the secure estimator at different degradation levels across varied SOC ranges for both charging and discharging of all packs. We use the data from these Monte Carlo runs to quantitatively analyze the algorithm's estimation accuracy, adaptability, and computational efficiency under the heuristic and GPR correction strategies. Table~\ref{tab:results_emp_gpr} summarizes our findings from these Monte Carlo runs for three degradation levels at $1^{st}$, $50^{th}$, and $100^{th}$ charging cycles of the battery. Fig.~\ref{fig:violin} further illustrates the distribution of estimation errors at these battery charging cycles with violin plots such that the spread indicates how frequently an error value occurs. The figure also includes box plots indicating the interquartile range along with the median, maximum, and minimum error lines.
  % \vspace{-6mm}
  \renewcommand{\arraystretch}{1.2}
  \begin{table}[h!]
      \centering
      \caption{Estimation performance comparison}
      \begin{tabular}{|c|l|c|c|}
      \hline
         Battery & \multirow{2}{*}{Parameters}  &  \multicolumn{2}{c|}{Methods} \\
         \cline{3-4}
        Cycle $\#$ &  & Heuristic & GPR \\
           \hline
           \hline
      \multirow{3}{*}{$1^{st}$ cycle}  & RMSE $[V]$  & 0.8357 & 0.4296 \\
         \cline{2-4}
         & Max Overestimate $[V]$ & 2.611 & 2.508 \\
          \cline{2-4}
        &  Max Underestimate $[V]$ & 1.419 & 1.489 \\
          \hline 
          \hline
          \multirow{3}{*}{$50^{th}$ cycle}  & RMSE $[V]$  & 0.8671 & 0.8718 \\
         \cline{2-4}
         & Max Overestimate $[V]$ & 1.471 & 1.981 \\
          \cline{2-4}
        &  Max Underestimate $[V]$ & 2.098 & 3.347 \\
          \hline 
          \hline
          \multirow{3}{*}{$100^{th}$ cycle}  & RMSE $[V]$  & 0.9646 & 1.8252 \\
         \cline{2-4}
         & Max Overestimate $[V]$ & 1.687 & 1.552 \\
          \cline{2-4}
        &  Max Underestimate $[V]$ & 2.678 & 5.063 \\
          \hline 
          \hline
      Overall  &  Run Time $[ms]$ & 0.112 & 0.293 \\
          \hline
      \end{tabular}
      \label{tab:results_emp_gpr}
  \end{table}
  %\noindent
  
  During testing for $1^{st}$ cycle charging of battery packs, the GPR correction outperforms the heuristic correction in terms of accuracy with a lower root-mean-squared error (RMSE) of $0.4296V$ compared to $0.8357V$ RMSE for the heuristic correction.  However, the GPR correction adapts poorly to changes in battery dynamics with aging and exhibits $103\%$ and $325\%$ escalations in RMSEs for the $50^{th}$ and $100^{th}$ cycle testing, respectively. In contrast, the heuristic correction demonstrates improved age adaptability with $4\%$ and $15\%$ increases in RMSEs for $50^{th}$ and $100^{th}$ cycle testings, respectively.  Fig.~\ref{fig:violin} shows that for heuristic correction, the widest error spreads and the median errors stay close to zero for all degradation-level testings. While the GPR correction yields tighter error bounds for new battery testing, both error dispersions and median errors significantly shift away from zero during the $50^{th}$ and $100^{th}$ cycle testing, as shown in Fig.~\ref{fig:violin}.
  
  %\noindent
  Lower values for the maximum over- and under-estimation of the voltage are crucial for efficient battery health management. Table~\ref{tab:results_emp_gpr} and Fig.~\ref{fig:violin} highlight that both methods exhibit an increasing trend in maximum underestimation of voltage with battery aging, while the maximum overestimation remains steady. The heuristic method exhibits a maximum underestimation of $2.678V$ with less than a 2-times rise over the $100^{th}$ cycle aging of the battery packs, while the GPR method exhibits $5.063V$ underestimation with more than 3.4 times rise under similar aging conditions.  
  Additionally, the heuristic correction achieves about $0.53$ times lower RMSE and maximum underestimation than the GPR correction during the $100^{th}$ cycle testing. This result illustrates the improved adaptability of the heuristic correction strategy.

  %\noindent
 In terms of resource efficiency, the GPR correction requires specific pre-trained models for each battery module, and thus runs with a much larger computational overhead. Conversely,  the heuristic correction utilizes small-data learning and limited cell-level knowledge captured by the $h (SOC, sgn({I}))$ function, and hence runs with minimal overhead.   Consequently, the heuristic correction exhibits a 2.63 times faster computational run-time compared to the GPR method and requires $0.112 ms$ on average to estimate one sample. Nevertheless, both methods can generate voltage estimates at a much faster rate of $3.4 kHz$ compared to a $ 1 Hz$ measurement sampling rate, and thus undermine the real-time applicability of the secure estimator.

  \subsection{Performance evaluation under compromised sensing}
  %\noindent
  In this section, we present two case studies to demonstrate the efficacy of the proposed self-learning Koopman-based secure estimator under compromised sensing, while considering different pack configurations, operating conditions, and prevalent cyberattack policies for each case study.

%\noindent
\textbf{Case study I: Denial-of-service during discharging:} 
\begin{figure}[h!]
    \centering
    \includegraphics[width=0.68\linewidth]{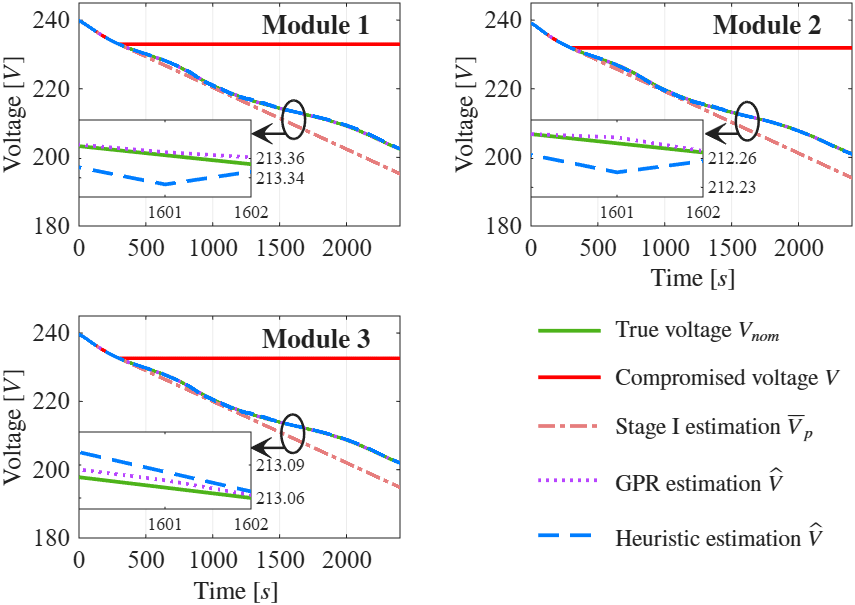}
    \caption{ \textbf{DoS sensor attack:} Each plot shows the true and compromised module voltage for the battery, the voltage estimation from self-learning Koopman-based estimation with only stage~I correction, with stage~II GPR correction, and with stage~II heuristic correction.  }
    \label{fig:dos}
\end{figure} 

In this scenario, we consider a Denial-of-service (DoS) sensor attack on the voltage measurement during discharging of the first battery pack.  Consequently, the cloud-BMS ceases to receive the updated voltage measurements, rather continues to work with the last module voltage measurement received before the DoS attack. Such DoS attacks may lead to over-discharging of the battery and may also adversely impact the grid. The attack starts at $300$s with a 0.819 battery SOC and continues for the next 35 minutes. This scenario is captured in Fig.~\ref{fig:dos} where the true and compromised module voltages are shown with green and red lines, respectively, in each plot. Under this DoS attack, the plots in  Fig.~\ref{fig:dos} show that the Stage~I correction performs poorly across all the modules. This phenomenon is expected as the Stage~I corrections are not equipped to incorporate higher-order complexities of battery dynamics and continue to use the predictive slope from the last correct learning window as the basis for prediction.  The plots in Fig.~\ref{fig:dos} show that both GPR and heuristic corrections estimate the module voltage with high accuracy (highlighted in the zoomed insets).  
%\noindent

\begin{figure}[h!]
    \centering
    \includegraphics[width=0.68\linewidth]{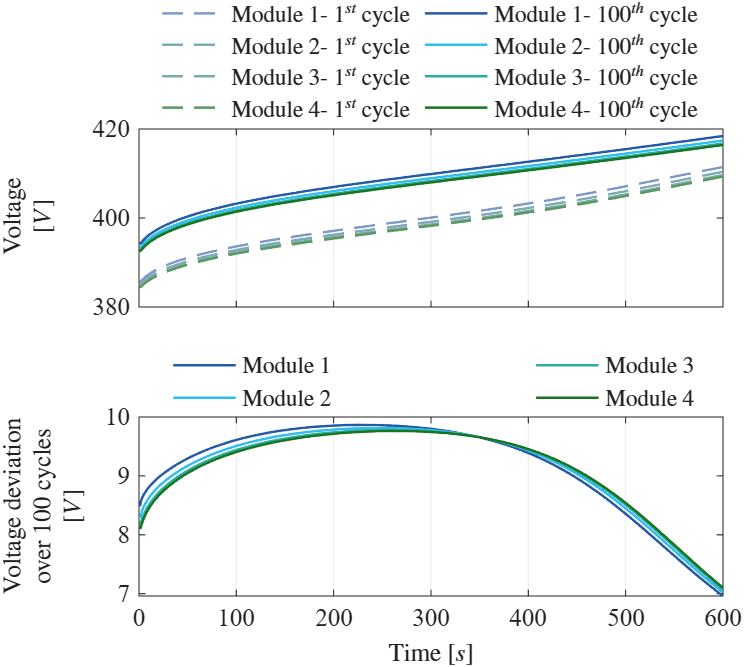}
    \caption{ \textbf{Battery aging: } The top plot shows the module voltage measurements of the battery pack during $1^{st}$ and $100^{th}$ charging cycles. The bottom plot illustrates the deviation in voltage measurements for each module over these 100 cycles of battery charging. }
    \label{fig:age_degrade}
\end{figure}

\textbf{Case study II: Bias attack on aging battery: } For this case study,  we consider an aging battery at $100^{th}$ charging cycle such that the battery capacity has degraded with aging. Consequently, under similar charging current inputs, the aged battery reaches a higher SOC level faster, leading to higher voltage measurements compared to a new battery. Fig.~\ref{fig:age_degrade} captures this deviation in module voltage measurements with aging, where top plot shows the module voltages measured at $1^{st}$ and $100^{th}$ charging cycles under similar charging conditions, and bottom plot illustrates the increment in voltage measurements due to battery degradation over 100 cycles. 

\begin{figure}[h!]
    \centering
    \includegraphics[width=0.68\linewidth]{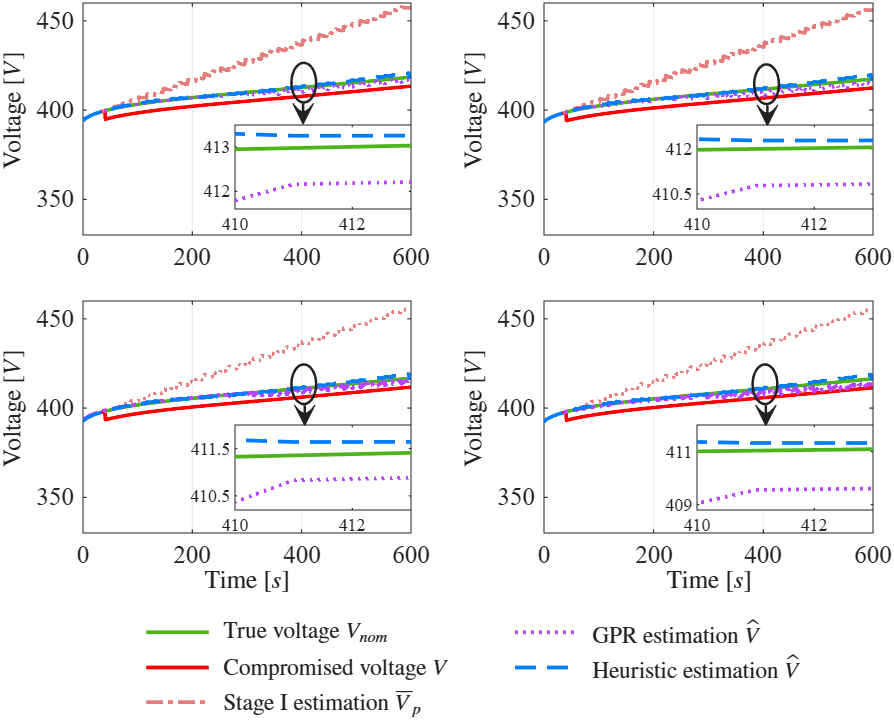}
    \caption{ \textbf{FDI sensor attack on aging battery: }Each plot shows the true and compromised module voltage for the battery, the voltage estimation from self-learning Koopman-based estimation with only stage~I correction, with stage~II GPR correction, and with stage~II heuristic correction. }
    \label{fig:bias_fdi}
\end{figure} 

%\noindent
 We then consider that the adversary introduces a false-data injection (FDI) sensor attack of negative $3V$ bias to all module voltage measurements at $40$s and continues to inject the bias attack for the remainder of the battery charging. Hence, after the first $40$s of charging, the cloud-BMS receives a $3V$ lower voltage measurement than the true $V_{nom}$ for all modules, which can lead to overcharging of the battery. The battery SOC at the start of the attack is 0.51. Fig.~\ref{fig:bias_fdi} captures this scenario, where each plot in Fig.~\ref{fig:bias_fdi} shows the true and compromised module voltages, respectively, with green and red lines.  The stage-I correction again fails to generate reliable module voltage estimations under this FDI attack, as shown in Fig.~\ref{fig:bias_fdi}. Moreover, the estimation error for the GPR correction increases with time due to the poor adaptability of the method. Conversely, the heuristic correction continues to generate highly accurate module voltage estimations for this aged battery, as highlighted in the zoomed insets in Fig.~\ref{fig:bias_fdi}. 
 
 Overall, both heuristic and GPR stage-II corrections provide reliable and accurate voltage estimations under compromised sensing for new batteries. Nevertheless, the GPR correction required individual model training for each battery pack; in contrast, the heuristic correction yielded comparable accuracy while using the same $h (SOC, sgn({I}))$ \eqref{total_compen} learned from a single LGM50 cell for both battery packs.  Additionally, during aged battery testing, the heuristic correction achieved a higher accuracy compared to the GPR correction.  These results illustrate the improved scalability as well as robustness towards battery degradation of the heuristic stage-II correction.

\section{Conclusion} \label{conclu}
In this work, we presented a self-learning Koopman operator-based secure voltage estimation algorithm for EVs under compromised sensing.
This self-learning Koopman operator uses its own prediction as feedback rather than corrupted measurements to evade the adverse impact of sensor attacks.
We also propose a two-stage error compensation strategy for the self-learning feedback to ensure accurate voltage estimation.  In stage~I error correction, we estimated the potential error in voltage prediction to compensate for the error accumulation from Koopman linear approximation. In stage~II, we proposed two methods to incorporate the higher-order battery characteristics into the baseline self-learning feedback.  For our first method, we propose a heuristic correction guided by cell-level OCV-SOC mapping that leverages this limited cell-level knowledge for module voltage estimation. 
Our second method uses a GPR-based data-driven error correction that circumvents the need for empirical derivation of the OCV-SOC map-guided heuristic function. We validate the proposed algorithm using high-fidelity simulation data under DoS and FDI sensor attacks during EV dis/charging for new and aged battery packs. Our simulation case studies illustrate that both heuristic and GPR corrections generate highly accurate voltage estimation under sensor attacks, while heuristic correction exhibits improved adaptability to battery aging.   Such secure estimation will enable the BMS to ensure optimal, safe, and coordinated  V2G charging of EVs while mitigating the deleterious impact of sensor cyberattacks.

\bibstyle{arxiv}
\bibliography{ref1,Ref_Troy}

\end{document}